%% file: __Incorrectness-POPL-TPSA-2025.tex
\pdfoutput=1
\documentclass[nonacm,sigconf,screen]{acmart} 
\usepackage{mathtools}
\usepackage{multirow}

\AtBeginDocument{%
  }

\setcopyright{none}

\usepackage{tikz}
\usetikzlibrary{patterns,decorations.pathreplacing,arrows,arrows.meta,decorations.pathmorphing,positioning,fit,trees,shapes,shadows,automata,calc,calligraphy,tikzmark}

\usepackage{xspace} 

\input{macros}

\begin{document}

\title{Partial Incorrectness Logic}

\author{Lena Verscht}
\affiliation{%
  \institution{RWTH Aachen}
  \country{Germany}
}
\affiliation{%
  \institution{Saarland University}
  \country{Germany}
}

\author{{\=A}nr{\'a}n W{\'a}ng}
\affiliation{%
  \institution{Saarland University}
  \country{Germany}
}

\author{Benjamin Lucien Kaminski}
\affiliation{%
  \institution{Saarland University}
  \country{Germany}
}
\affiliation{%
  \institution{University College London}
  \country{United Kingdom}
}

\renewcommand{\shortauthors}{Verscht, W{\'a}ng, Kaminski}

\maketitle


\section{Introduction}
\label{sec:intro}
Reasoning about program correctness has been a central topic in static analysis for many years,  with \emph{Hoare logic} (HL) playing an important role~\cite{hoareAxiomaticBasisComputer1969, aptFiftyyearsHoares2019}. 
The key notions in HL are \emph{partial} and \emph{total correctness}.
Both require that program executions starting in a specified set of initial states (the \emph{precondition}) reach a designated set of final states (the \emph{postcondition}).
Partial correctness is more lenient in that it does not require termination, effectively deeming divergence acceptable.

We explore \emph{partial incorrectness logic} \cite{zhangQuantitativeStrongestPost2022}, which stands in relation to O'Hearn's \enquote{total} incorrectness logic \cite{ohearnIncorrectnesslogic2020} as partial correctness does to total correctness:
Partial correctness allows divergence, partial \emph{incorrectness} allows \emph{unreachability}.
While the duality between divergence and unreachability may not be immediately apparent, we explore this relationship further.

Our chosen formalism is \emph{predicate transformers} à la Dijkstra~\cite{dijkstraGuardedcommandsnondeterminacy1975}.
We focus here on \emph{deterministic} and \emph{reversible} programs, though the discussion extends to nondeterministic and irreversible computations, both of which introduce additional nondeterminism that must be addressed.


%

\section{Correctness and Termination}
\label{sec:correctness}
Hoare logic \cite{hoareAxiomaticBasisComputer1969} is concerned with triples of the form $\hoare{\pre}{\program}{\post}$, where~$\pre$ is a precondition on initial states, $\program$ is a program, and $\post$ is a postcondition on final states.
$\hoare{\pre}{\program}{\post}$ is said to be \emph{valid for total correctness} if all computations of $\program$ started in initial states satisfying precondition $\pre$ (1) terminate, and (2) do so in some state satisfying the postcondition $\post$.

As is well known, proving termination is undecidable.
This is motivation to define a more lenient notion of correctness, \emph{partial correctness}, which requires only criterion (2) above.
In other words: \emph{If} $\program$ terminates, it has to do so in $\post$.

\subsection{Weakest (Liberal) Pre}
\label{ssec:weakestpre}
We use \emph{predicate transformers} due to Dijkstra as the formal framework for expressing program specifications \cite{dijkstraPredicateCalculusProgram1990}.
The \emph{weakest precondition} $\wp{\program}{\post}$ of a postcondition $\post$ with respect to a program~$\program$ contains \emph{all} states from which $\program$ always terminates in $\post$.
Notably, this is the weakest predicate $\pre$ such that $\hoare{\pre}{\program}{\post}$ is valid for \emph{total} correctness.
Dually, the \emph{weakest liberal precondition} $\wlp{\program}{\post}$ of $\post$ with respect to~$\program$ contains \emph{all} states from which $\program$ either diverges or terminates in $\post$.
This is the weakest predicate $\pre$ such that $\hoare{\pre}{\program}{\post}$ is valid for \emph{partial} correctness.

Computing weakest (liberal) preconditions can be done inductively on the program structure.
The most intricate part of this are loops, which require reasoning about fixed points.
The weakest (liberal) preconditions of while loops are defined as fixed points of a characteristic function $\Phi$ which, intuitively, captures one execution of the loop body:
%
%
\begin{align*}
    \wp{\WHILEDO\varphi{C'}}{F} &\eeq \mu\,\Phi\\
    \text{and \qquad } \wlp{\WHILEDO\varphi{C'}}{F} &\eeq \nu\,\Phi~,
\end{align*}%
\noindent%
where $\mu\,\Phi$ denotes the least and $\nu\,\Phi$ the greatest fixed point of $\Phi$~\cite{kaminskiAdvancedWeakestPrecondition2019}.
For our intents, the relevant aspect here is that weakest preconditions are \emph{least} fixed points while weakest liberal preconditions are \emph{greatest} fixed points.

\subsection{Total and Partial Correctness}
\label{ssec:partialtotalcorr}
Above, we have hinted at the close connection between total (partial) correctness and weakest (liberal) preconditions, which comes as an \emph{underapproximation}:

\noindent
\hspace{-5pt}
%
\begin{tabular}{l|ll}
\multirow{2}{*}{$\hoare{\pre}{\program}{\post}$ is valid for}
     & total correctness \quad &iff \quad $\pre \subseteq \wp{\program}{\post}$ \\
     & partial correctness \quad &iff \quad $\pre \subseteq \wlp{\program}{\post}$ \\
\end{tabular}

For proving total correctness of a loop, we thus have to prove a {lower} bound on a \emph{least} fixed point -- 
for partial correctness, we also have to prove a {lower} bound, but on a \emph{greatest} fixed point.
The latter can be conveniently proven using the following theorem:
%
\begin{lemma}[Park Induction]\label{lem:park}
    Let $(X, {\preceq})$ be a complete partial order and $f\colon X \to X$ be monotonic.
    Then for all $I \in X$,%
    \begin{align*}
        & I\ppreceq f(I) \qimplies I \ppreceq \nu f.
    \end{align*}%
    %
\end{lemma}%
\noindent%
If we have a suitable candidate or \emph{invariant} $I$, we can thus prove \emph{lower} bounds on \emph{greatest} 
fixed points by just one application of $f$.
For \emph{lower} bounds on \emph{least} fixed points, there is no such simple rule.
The consequence for proofs of correctness is that we can prove partial correctness easily, while total correctness requires more intricate arguments~\cite{harkAimingLowHarder2020}. 


\subsection{Termination}
\label{ssec:termination}
Given a proof of partial correctness, we prove total correctness by additionally proving termination.
In terms of weakest preconditions, this means
\begin{align*}
    \underbrace{\pre \subseteq \wp{\program}{\post}}_{\text{\scriptsize total correctness}} \quad \impliedby \quad \underbrace{\pre \subseteq \wlp{\program}{\post}}_{\text{\scriptsize partial correctness}} \, \land \, \underbrace{\pre \subseteq \wp{\program}{\true}}_{\text{\scriptsize termination}}.
\end{align*}%
Note that we in fact do not require \emph{universal} termination, i.e.\ termination from \emph{all} initial states, but only from the states of interest (those satisfying $\pre$).

There are several reasons why to separate these concerns:
In the spirit of divide and conquer, we can split a problem into smaller subproblems.
Proving partial correctness is indeed conceptually easier than proving total correctness, as we have seen in {\color{ACMPurple}Section}~\ref{ssec:partialtotalcorr}.
Additionally, proving termination is a subfield on its own, meaning we can profit from well developed techniques and algorithms.


\paragraph{Proving termination using variants}
To prove universal termination
of a while loop $\WHILEDO{\varphi}{\program}$, we can make use of \emph{variants}:
A variant is a function $\variant \colon \states \to \Nats$ such that for all $n \in \Nats$, the triple $\hoare{\varphi\wedge v=n}{\program}{ v<n}$ is valid for total correctness. 
Intuitively, each program state is assigned a natural number that has to strictly decrease in every iteration.
Since the natural numbers are well-founded, there cannot be infinitely many iterations.
Therefore, if a variant exists, then the loop $\WHILEDO{\varphi}{\program}$ terminates. 

\section{Incorrectness and Reachability}
\label{sec:incorrectness}

Reasoning about program correctness is a well-established area of research.
Recently, a dual concept has gained significant attention: program \emph{incorrectness} \cite{ohearnIncorrectnesslogic2020,devriesReverseHoareLogic2011}.
Intuitively, incorrectness is concerned with reachability of a set of final states assumed to represent unwanted behavior, or, \emph{bugs}.
A triple $\incorrectness{\pre}{\program}{\post}$ is valid for incorrectness if all states satisfying $\post$ are reachable via execution of program $\program$ from some initial state in $\pre$.

Similar to correctness, we can also characterize incorrectness as the underapproximation of a predicate transformer.
This raises the question whether for incorrectness we can have a similar separation of concerns as with total correctness in the form of
\begin{align*}
    \text{incorrectness} \eeq \text{something} + \text{something else}~.
\end{align*}%
We will see that we can answer this positively.
For this, let us again start by looking at the required predicate transformers.

\subsection{Strongest (Liberal) Post}
The \emph{strongest postcondition} $\sp{\program}{\pre}$ of a program $\program$ and a precondition $\pre$ contains all \emph{final} states which computation starting from $\pre$ can reach~\cite{dijkstraPredicateCalculusProgram1990}.
%
Strongest \emph{liberal} postconditions on the other hand are much less investigated and were only recently introduced by Zhang and Kaminski~\cite{zhangQuantitativeStrongestPost2022}.
Liberality here refers to permitting unreachability, in contrast to divergence in the correctness setting.
This comes rather natural, because predicate transformers, as other approaches, take a relational view on computation:
In the end, we are not so much interested in what happens within the program, instead, we abstract this to which initial states lead to (or \emph{are related to}) which final states.

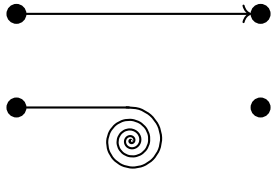
\begin{figure}
    \centering
    \begin{tikzpicture}[->,
				node distance=1cm and 3cm,
				thick
				]

				\node[state,fill=black,scale=0.3] (I1) {\phantom{1}};
				\node[state,fill=black,scale=0.3] (F1) [right =of I1] {\phantom{1}};
                \node[state,fill=black,scale=0.3] (I2) [below =of I1] {\phantom{1}};
				\node[state,fill=black,scale=0.3] (F2) [right =of I2] {\phantom{1}};

				\node (spiral) at ($(I2)!0.5!(F2)$) {};

				\path (I1) edge node {} (F1);
				\path (I2) edge [-] node {} (spiral);

				\coordinate (spiralstart) at ($(spiral)+(-0.1,-0.442)$);
				\draw let \p1=(I2) in
				[scale=0.25,domain=0:30,variable=\t,smooth,samples=75,-,shift={(spiralstart)}] plot  ({\t r}: {-0.002*\t*\t});
			
			\end{tikzpicture}
    \caption{Termination versus reachability.}
    \label{fig:simple-program}
\end{figure}

A simple illustration for this is provided in \autoref{fig:simple-program}.
The program state space only spans two states.
Starting in the first state the program terminates, from the second state the program diverges.
We see that the second initial state from which computation diverges is related to no final state.
Dually, the second final state is related to no initial state; it is \emph{unreachable}. 
Therefore, the strongest liberal post $\slp{\program}{\pre}$ contains the states which computation starting from $\pre$ can reach, and additionally the \emph{unreachable states}.

Similar to weakest (liberal) preconditions, the strongest (liberal) postconditions can be computed using inductive rules.
Those for the $\slpsymbol$ calculus with predicates are given in \autoref{tab:slp-def}. 

The loop definitions again invoke \emph{least} and \emph{greatest} fixed points of a characteristic function $\Phi$:%
\begin{align*}
    &\sp{\WHILEDO\varphi{C'}}{F} = \mu\,\Phi\\ 
    \text{and \qquad } &\slp{\WHILEDO\varphi{C'}}{F} = \nu\,\Phi.
\end{align*}
Again, the liberal calculus corresponds to a \emph{greatest} fixed point and the non-liberal to a \emph{least} fixed point.
%

\begin{table}
  \renewcommand{\arraystretch}{1.3}
  \begin{tabular}{ll}
    \toprule
    $C$&$\slp{\program}{F}$\\
    \midrule
    $\DIVERGE$ & $\true$\\
    $\ASSIGN{x}{e}$ & $\forall\alpha:x\neq e[x/\alpha]\vee F[x/\alpha]$ \\
    $\COMPOSE{C_1}{C_2}$ & $\slp{C_2}{\slp{C_1}{F}}$\\
    $\ITE{\varphi}{C_1}{C_2}$ & $\slp{C_1}{\neg\varphi\vee F}\wedge \slp{C_2}{\varphi\vee F}$\\
    $\WHILEDO{\varphi}{C'}$ & $\varphi\vee\big(\nu Y.\,F\wedge\slp{C'}{\neg\varphi\vee Y}\big)$ \\
  \bottomrule
\end{tabular}
  \caption{The strongest liberal postcondition transformer~\cite{zhangQuantitativeStrongestPost2022}}
  \label{tab:slp-def}
\end{table}

\subsection{Total and Partial Incorrectness}
Similar to total correctness, we can characterize \emph{in}correctness as an underapproximation of the \emph{non-liberal} strongest post.
It is thus a natural choice to define a partial variant of incorrectness as an underapproximation of the \emph{liberal} transformer, giving us:

\noindent
\hspace{-5pt}
\begin{tabular}{l|ll}
\multirow{2}{*}{$\incorrectness{\pre}{\program}{\post}$ is valid for}
     & \enquote{total} incorrectness  &iff \ \  $\post \subseteq \sp{\program}{\pre}$ \\
     & partial incorrectness  &iff \ \  $\post \subseteq \slp{\program}{\pre}$ \\
\end{tabular}


For partial incorrectness, $\post$ may contain unreachable states, but \emph{if} a state satisfying $\post$ is reachable, then it was reachable from some initial state satisfying $\pre$.

Using Park induction ({\color{ACMPurple}Lemma}~\ref{lem:park}) again, we can prove partial incorrectness easily (upon finding a suitable invariant).
Yet again, proving total incorrectness by finding a lower bound on the non-liberal $\spsymbol$ is hard, which is why we divide this problem into subproblems in the next section.

\subsection{Reachability} \label{ssec:reachability}
We can derive partial from total correctness by discharging termination. 
Dually, we obtain partial incorrectness from \enquote{total} incorrectness by discharging reachability.
Reachability is expressible in terms of predicate transformers, very similar as to what we have seen for termination:
\[
    \underbrace{\post \subseteq \sp{\program}{\pre}}_{\text{\scriptsize \enquote{total} incorrectness}} \quad \impliedby \quad \underbrace{\post \subseteq \slp{\program}{\pre}}_{\text{\scriptsize partial incorrectness}} \, \land \, \underbrace{\post \subseteq \sp{\program}{\true}}_{\text{\scriptsize reachability}}.
\]
The strongest post of $\true$ contains all final states that are reachable.
Again, we do not have to require reachability of \emph{all} final states, but only of those in the postcondition $\post$.

\paragraph{Proving reachability using variants}
Unreachability is caused by more than loops: 
Even a simple assignment can already introduce unreachable states.
Still, loops remain the most complex part of the reachability proof, as it requires the computation of a fixed point.
Dual to the variant method to prove termination, there are (backward) variant rules to prove reachability~\cite{ohearnIncorrectnesslogic2020, devriesReverseHoareLogic2011}. 
A function $\variant \colon \states \to \Nats$ is called a \emph{backward variant} if for all $n \in \Nats$, the triple $\incorrectness{v<n \land \varphi}{\program}{v=n}$ is valid for incorrectness.
Intuitively, each state in which the variant has value $n$ must be reachable from a state with value strictly less than $n$.
Again, since the natural numbers are well-founded, there are only finitely many values strictly less than $n$.
Therefore, if a backwards variant $\variant$ exists, then there exists an $n \in \Nats$ such that the states where $\neg\varphi$ holds and $\variant=n$ are reachable when  executing $\WHILEDO{\varphi}{\program}$.

Note that this does not prove \emph{universal} reachability, i.e.\ reachability of \emph{all} final states, but only of those where the loop guard does not hold and where the variant takes on value $n$. 
It would be interesting to see if we can find rules that prove reachability of the states in some postcondition $\post$ only.

\subsection{Partial Incorrectness: Applications}
Beyond serving as a step toward classic (\enquote{total}) incorrectness, partial incorrectness is an interesting property on its own.
We now explore two scenarios where partial incorrectness proves useful.

\paragraph{Underapproximation triples}
Consider the following program inspired by the beginning example of \cite[Section 3.1]{ohearnIncorrectnesslogic2020}:
\[
    p\coloneqq \ITE{x \text{ even}}{\ASSIGN{y}{y+1}}{\ASSIGN{y}{2\cdot y}}
\]
%
Assume the precondition $y=10$ and the postcondition $y=11$.
As \citet{ohearnIncorrectnesslogic2020}, we can now ask:
Is this postcondition an under-approximation of the states reachable by executing $P_y$ starting from the precondition. 
Or equivalently:
Is triple $\incorrectness{y=10}{p}{y=11}$ valid for incorrectness?

It is not.
This seems unintuitive at first.
However, consider a final state where $y=11$ and $x=1$.
This state is unreachable from initial states where $y=10$; in fact, it is entirely unreachable from \emph{any} initial state.
This is because if $x$ is odd, as it is in this example, the final value of $y$ must be even.

This is where \emph{partial} incorrectness logic comes into application.
The triple $\incorrectness{y=10}{p}{y=11}$ \emph{is} valid for partial incorrectness!
Similar to how partial correctness treats divergence as acceptable behaviour, partial incorrectness accepts universal unreachability.
By doing so, we mitigate the cause of the unintuitive scenario encountered before.
\begin{figure}
    \centering
    \begin{align*}
    & \annotate{y=10} \\
    & \IFSYMBOL\,(\, {x \text{ even }} \, )\,\{\, \\
    & \quad \annotate{x \textbf{ odd} \lor y=10} \\
    & \quad \ASSIGN{y}{y+1}  \\
    & \quad \annotate{x \textbf{ odd} \lor  y=11} \\
    & \}\,\ELSESYMBOL\, \{\\
    & \quad \annotate{x \textbf{ even} \lor  y=10} \\
    & \quad \ASSIGN{y}{2 \cdot y} \\
    & \quad \annotate{x \textbf{ even} \lor y \textbf{ odd} \lor y=20} \\
    & \} \\
    & \annotate{(x \textbf{ odd} \lor  y=11) \land (x \textbf{ even} \lor y \textbf{ odd} \lor y=20)}
\end{align*}
\caption{Calculating the strongest liberal post.}
    \label{fig:calc}
\end{figure}


Formally, we can prove the validity of the triple for partial correctness by calculating $\slp{p}{y=10}$, which is done by applying the rules presented in \autoref{tab:slp-def}.
Detailed calculations are shown in \autoref{fig:calc}.
Since the postcondition $y=11$ implies $\slp{p}{y=10} = (x \text{ odd } \lor  y=11) \land (x \text{ even } \lor y \text{ odd} \lor y=20)$, the above triple is valid for partial incorrectness.
Thus, partial incorrectness can help focus the reasoning on variables of interest.



\paragraph{Backtracking responsibilities} 
With partial incorrectness triples in the form of $\incorrectness{\_}{\program}{\text{true}}$, we can reason about responsibilities:
The blank can be filled by initial states that are responsible for some result.
Consider Schrödinger's cat sitting in a sealed box, described by the program $\program:= \WHILEDO{\neg \open}{\ASSIGN{\dead}{\spill}};\  \ASSIGN{\dead}{\spill}$. 
The cat may or may not be $\dead$ depending on whether a vial of poison was $\spill$ed. 
Once we $\open$ the box, we observe the cat's health. 
If we do not open the box, the variable $\dead$ may be true or false at any time, but we have no observations. 

Here, $\incorrectness{\open}{\program}{\true}$ is a valid partial incorrectness triple. 
The postcondition $\true$ tells us that any reachable final state must have an origin satisfying $\open$:
The action of opening the box caused us to have knowledge over the cat. 

Similar reasoning can be acquired by \emph{total} incorrectness.
However, this is subject to the strong condition that there are no unreachable states at all.
For this example, $\incorrectness{\text{open}}{\program}{\text{true}}$ is indeed invalid for total incorrectness.

\section{Conclusion}
\label{sec:conclusion}
We discussed the relatively unexplored notion of partial incorrectness.
In doing so, we stressed the duality between the weakest (liberal) pre and strongest (liberal) post; correctness and incorrectness; termination and reachability.
We argue that partial incorrectness also proves useful in computational terms, as does partial correctness: 
Park's induction allows simple proofs of \emph{partial} incorrectness but not of \emph{total} incorrectness.
By adding a proof of reachability, we can get from partial to total incorrectness.

The examples show that partial incorrectness allows us to drop irrelevant variables and reason about responsibilities.
Further applications for partial incorrectness are still open to investigation.



\paragraph{Quantitative programs}
For simplicity, we chose the pre and post to be Boolean predicates. 
However, techniques mentioned here extend to quantitative \emph{expectations} \cite{morgan1996probabilistic,kaminskiAdvancedWeakestPrecondition2019} and we conjecture that the results transfer.

\paragraph{Proving reachability}
We presented a variant rule for proving reachability of loops.
However, assignments can also cause unreachability, and their computation is not straightforward; efficient methods for this would be valuable to investigate.

\paragraph{Nomenclature}
We feel that the names ``correctness'' and ``incorrectness'' are not covering the entire range of applications of the logics. 
Certainly, when we choose the postcondition to represent bugs, we reason about unwanted properties of our programs, but as is shown in our examples, we can just as well analyse properties where the post represents desirable final states.





\bibliographystyle{ACM-Reference-Format}
\bibliography{__references}


\end{document}

%% file: macros.tex
\usepackage{stmaryrd}
\usepackage{csquotes}

\newcommand{\states}{\ensuremath{\text{States}}\xspace}

\newcommand{\incorrectness}[3]{\left[ \,{#1}\vphantom{#3}\, \right] \mathrel{#2} \left[ \, {#3}\vphantom{#1} \, \right]}
\newcommand{\hoare}[3]{\left\langle \,{#1}\vphantom{#3}\, \right\rangle \mathrel{#2} \left\langle \, {#3}\vphantom{#1} \, \right\rangle}

\newcommand{\sfsymbol}[1]{\textsf{\upshape {#1}}}

\newcommand{\wpsymbol}{\sfsymbol{wp}}

\renewcommand{\wp}[2]{\wpsymbol\,\llbracket#1\rrbracket(#2)}

\newcommand{\spsymbol}{\sfsymbol{sp}}

\renewcommand{\sp}[2]{\spsymbol\,\llbracket#1\rrbracket(#2)}

\newcommand{\slpsymbol}{\sfsymbol{slp}}

\newcommand{\slp}[2]{\slpsymbol\llbracket#1\rrbracket(#2)}

\newcommand{\wlpsymbol}{\sfsymbol{wlp}}

\newcommand{\wlp}[2]{\wlpsymbol\llbracket#1\rrbracket(#2)}

\newcommand{\conditionalPair}[2]{{\let\oldarraystretch\arraystretch}\renewcommand{\arraystretch}{1}~\holter{~\raisebox{.5ex}{${#1}$}~}{~\raisebox{.125ex}{${#2}$}~}~\renewcommand{\arraystretch}{\oldarraystretch}}

\newcommand{\annotate}[1]{\boldsymbol{\annocolor{\!\fatslash\!\!\!\fatslash~~\vphantom{G'} {#1}}}}

\newcommand{\DIVERGE}{\ensuremath{\textnormal{\texttt{diverge}}}}

\newcommand{\AssignSymbol}{\coloneqq}
\newcommand{\ASSIGN}[2]{\ensuremath{#1 \AssignSymbol #2}}

\newcommand{\AVAILLOC}[1]{\PosNats}
\usepackage{actuarialangle}

\newcommand{\COMPOSE}[2]{\ensuremath{{#1}{\,\fatsemi}~ {#2}}}

\newcommand{\IFSYMBOL}{\ensuremath{\textnormal{\texttt{if}}}}

\newcommand{\ELSESYMBOL}{\ensuremath{\textnormal{\texttt{else}}}}

\newcommand{\ITE}[3]{\ensuremath{\IFSYMBOL\,\left(\, {#1} \,\right)\,\left\{\, {#2} \,\right\}\,\ELSESYMBOL\,\left\{\, {#3} \,\right\}}}

\newcommand{\WHILESYMBOL}{\ensuremath{\textnormal{\texttt{while}}}}

\newcommand{\WHILEDO}[2]{\ensuremath{\WHILESYMBOL \left(\, {#1} \,\right)\left\{\, {#2} \,\right\}}}

\newcommand{\Nats}{\ensuremath{\mathbb{N}}\xspace}
\newcommand{\PosNats}{\ensuremath{\mathbb{N}_{>0}}\xspace}

\newcommand{\true}{\mathsf{true}}

\newcommand{\dead}{\mathsf{dead}}
\newcommand{\spill}{\mathsf{spill}}
\newcommand{\open}{\mathsf{open}}

\newcommand*\oline[1]{%
  \vbox{%
    \hrule height 0.3pt
    \kern0.25ex
    \hbox{%
      \kern0.04em
      \ifmmode#1\else\ensuremath{#1}\fi
      \kern0.04em
    }
  }
}

\newcommand*\olinedbl[1]{%
  \vbox{%
    \hrule height 0.6pt
    \kern0.25ex
    \hbox{%
      \kern0.04em
      \ifmmode#1\else\ensuremath{#1}\fi
      \kern0.04em
    }
  }
}

\newcommand{\qimplies}{\quad\textnormal{implies}\quad}

\newcommand{\ppreceq}{~{}\preceq{}~}

\newcommand{\eeq}{~{}={}~}

\definecolor{webgreen}{rgb}{0,.5,0}

\newcommand{\annocolor}[1]{\textcolor{webgreen}{#1}}

\newcounter{computationarrowsone}
\newcounter{computationarrowstwo}

\newcounter{sarrow}

\newcommand{\pre}{\ensuremath{b}}
\newcommand{\post}{\ensuremath{c}}
\newcommand{\program}{\ensuremath{p}}

\newcommand{\variant}{\ensuremath{v}}

\definecolor{reeed}{rgb}{0.59, 0.0, 0.09}
